\documentclass[12pt,preprint]{aastex}

\begin{document}

\title{Hubble Space Telescope Observations of Star Clusters in M101%
\footnote{
Based on observations made with the NASA/ESA Hubble Space Telescope, obtained at the Space 
Telescope Science Institute, which is operated by the Association of Universities for Research 
in Astronomy, Inc., under NASA contract NAS 5-26555. These observations are associated with 
programs \#8640 and \#9490.}
}

\author{Pauline Barmby}
\affil{Harvard-Smithsonian Center for Astrophysics, 60 Garden St., Mailstop 65,
Cambridge, MA 02138}
\email{pbarmby@cfa.harvard.edu}

\author{K.D. Kuntz}
\affil{
The Henry A. Rowland Department of Physics and Astronomy,
The Johns Hopkins University, 400 Charles Street, Baltimore MD 21218
and Exploration of the Universe Division, 
X-Ray Astrophysics Laboratory, Code 662, 
NASA Goddard Space Flight Center, Greenbelt, MD 20771
}

\author{John P. Huchra}
\affil{Harvard-Smithsonian Center for Astrophysics, 60 Garden St.,
Cambridge, MA 02138}

\author{Jean P. Brodie}
\affil{UCO/Lick Observatory, University of California, Santa Cruz, CA 95064}

\begin{abstract}
{\it Hubble Space Telescope} Advanced Camera for Surveys (ACS)
images are used to identify and study star cluster candidates in the
nearby spiral galaxy M101.
About 3000 round, slightly-resolved cluster candidates are identified in 
10 ACS pointings covering an area of 106~arcmin$^2$. The cluster candidates'
color and size distributions are consistent with those of star clusters 
in other nearby spirals.
The majority of the M101 candidates are blue and more likely to be
associated with the galaxy's spiral arms, implying that they
are young. The galaxy-luminosity-normalized number of `young massive clusters'
in M101 is similar to that found in other spirals,
as is the cluster density at a fiducial absolute magnitude.
We confirm a previous finding that M101 has a large number of faint, 
red star clusters: if these are old globular clusters then this galaxy
has a very large globular cluster population. More plausible 
is that the faint red clusters are reddened young clusters; their
colors and luminosities are also consistent with this explanation.
\end{abstract}

\keywords{galaxies: individual (M101) --- galaxies: star clusters  --- galaxies:spiral}

\section{Introduction}

Understanding the stellar populations of galaxies is key to
untangling the mysteries of their formation and evolution.
Because of our location in the Milky Way, getting an overall
picture of the stellar populations is difficult.
The nearby spiral galaxy  M101 (NGC~5457) is the closest face-on 
spiral \citep[Hubble type SAB(rs)cd;][]{rc3}  and 
provides an excellent opportunity for resolution of
stellar population details. Previous studies of stellar populations
in M101 include work on the Cepheid variable stars \citep{kel96,ste98},
X--ray binaries \citep{kdk05,mukai03}, and novae \citep{scp00},
among many others.

Star clusters are excellent tracers of stellar populations:
they are much brighter than single stars and usually
have small internal spreads in age and metallicity. Globular clusters
in particular are believed to be indicators of galaxy 
history \citep[e.g.,][]{str05}. Globular cluster systems (GCSs) 
in spirals have been studied much less than
their counterparts in ellipticals: spirals have fewer clusters per unit
of galaxy light, and their irregular background light 
makes detection of clusters more challenging.
The Milky Way GCS is the `gold standard' for comparison to all others, 
but it is difficult to say whether it is a truly typical GCS
since the sample of spiral galaxy GCSs available for comparison is small.
Younger star cluster populations are better-studied in other galaxies
\citep[e.g.,][]{lr99} than in the Milky Way; due to extinction, the census of 
young cluster populations in the Galaxy is far from complete \citep{bica03}.

Studying clusters in edge-on spirals \citep[e.g.,][]{pg03} ameliorates the background 
light issue, but the nearest such galaxies are distant enough that their star
clusters are unresolved; confirmation that candidates are true clusters is difficult.
Recently, {\it Hubble Space Telescope} (HST) imaging has been used to
search for star clusters in a number of low-inclination nearby spiral galaxies,
including M81 \citep[Nantais et al., 2006 in preparation]{cft1}, 
M51 \citep{bik03,lcw05}, and these two galaxies plus
M83, NGC~6946 and M101 \citep{cwl04}.
The high spatial resolution of HST images aids in 
distinguishing true clusters from contaminants such as 
background galaxies and multiple star blends and improves
photometric precision. 
The HST studies mentioned above have generally covered only small regions
of these nearby galaxies using one or two HST pointings. Here we complement
these works by presenting initial results from a study of star clusters found in 
ten HST/ACS fields in M101, covering a much larger fraction of the galaxy.
Throughout this work we assume a distance to M101 of 6.7~Mpc \citep[$m-M=29.13$;][]{hst_kp}.
At this distance, one arcsecond subtends 32.5~pc.

\section{Observations and cluster selection}

\begin{figure}
\plotone{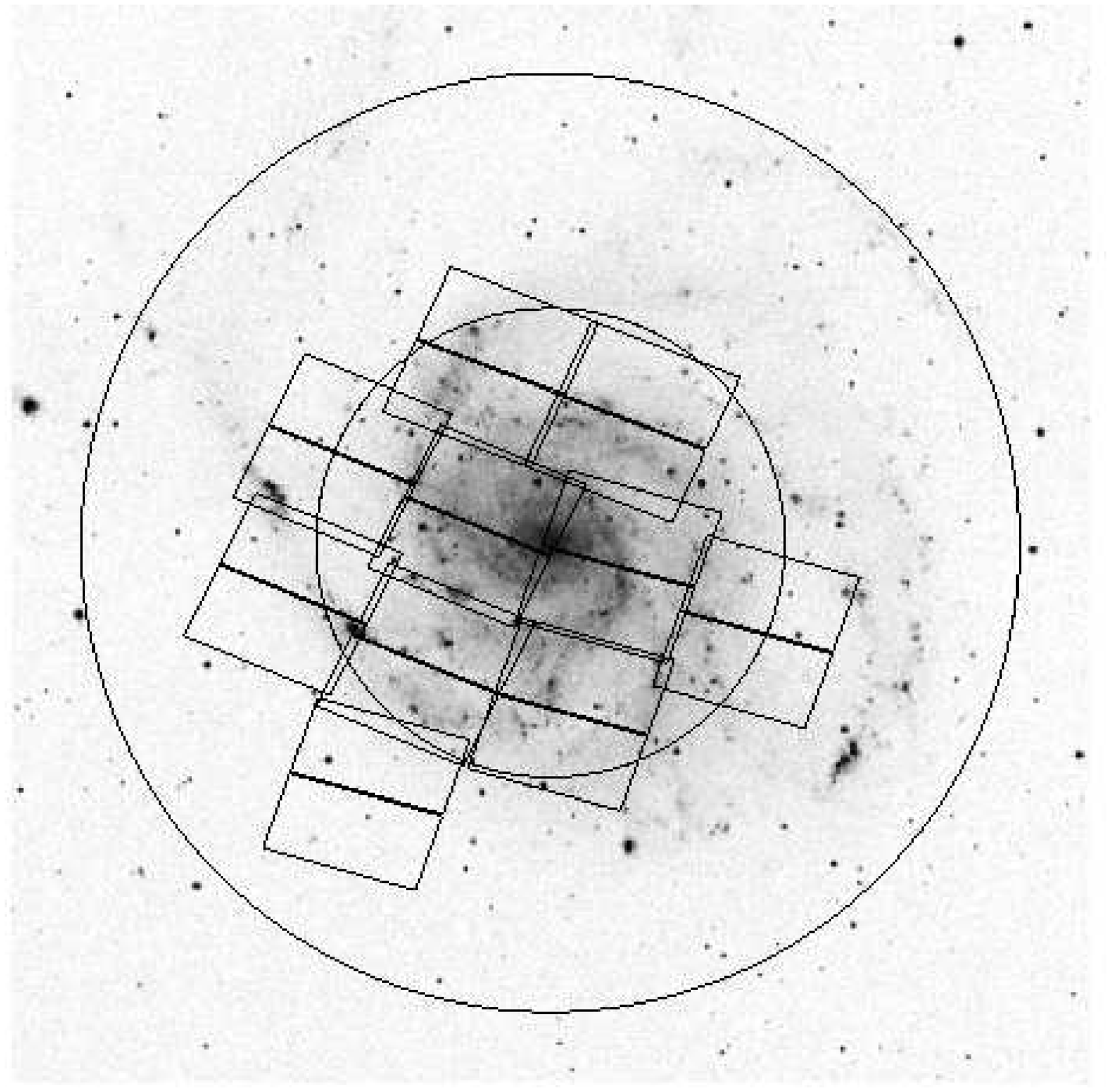}
\caption{Location of ACS fields in M101, overplotted on a Digitized
Sky Survey image. North is up and East to the left.
The two circles are centered on the galaxy and 
have radii of 5 and 10 arcminutes, or 9.75 and 19.5~kpc
at the distance of M101 \citep[6.7~Mpc;][]{hst_kp}.
\label{m101-fields}}
\end{figure}

Images of 10 fields in M101 (see Figure~\ref{m101-fields}) were taken with the
Wide Field Camera of the Advanced Camera for Surveys \citep{acs} on HST, during the period  
2002 November 13--16. Each field was observed for 900~s in the F435W filter and 
720~s each in F555W and F814W; the exposures were `cosmic-ray-split'
but not dithered. The pipeline-processed, drizzled 
versions of the data retrieved in 2003 February
were used in this analysis. The $S/N=10$ detection limit for point sources
in these data is given by the ACS exposure time calculator as
$B=25.7$, $V=25.4$, $I=25.0$. The total area covered by our survey
is about 106~arcmin$^2$; this accounts for the overlap between
the ACS pointings and the missing area in the ACS inter-chip gaps.

\begin{figure}
\plotone{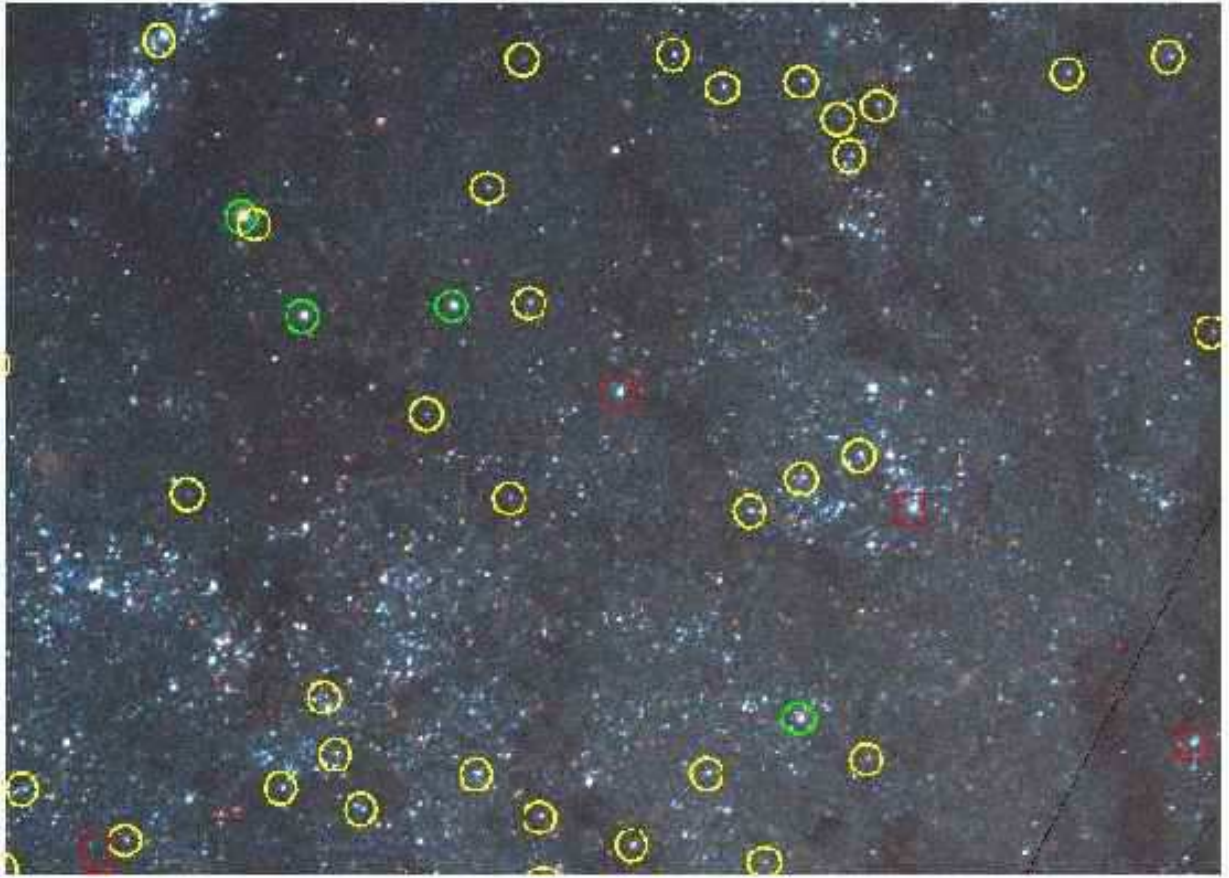}
\caption{Three-color view of a portion (1\farcm0$\times$0\farcm7) of
one ACS field, centered at position 
$14^{\mathrm h}03^{\mathrm m}27$, $54\arcdeg22\arcmin13\arcsec$ (J2000).
North is up and east to the left.
The F435W filter is blue, F555W is green, F814W is red.
Yellow circles are cluster candidates selected from the ACS data,
green circles are cluster candidates selected from the ACS data
and by \citet{bks96}, and red squares mark candidates selected by
\citet{bks96} but not in the present work.
\label{m101-detail}}
\end{figure}

Even a casual inspection of the images reveals a wealth of detail
(e.g., see Figure~\ref{m101-detail}), and an initial attempt at source extraction
yielded tens of thousands of detected objects per field. The expected
population of M101 star clusters includes a few hundred globular clusters and 
a few thousand younger clusters, so finding these 
requires care and attention. Although automatic selection can reduce
the number of  cluster candidates to a manageable one, we found that
human judgment was required in the final selection.

Star clusters in M101 were expected to be slightly resolved
in HST images: a typical Milky Way globular cluster half-light radius of 
3.2 pc corresponds to about 2 ACS/WFC pixels. Unfortunately there are many other 
sources which are slightly resolved in the images, including
background galaxies, H~II regions, blended stars, etc.
To understand what star clusters in M101 might look like,
we modeled their appearance using HST images of M31 globular clusters.
The M31 cluster images were taken with STIS as part of program
GO-8640, designed to measure structural parameters for the clusters.
We chose six clusters with a range of sizes and shapes, and
modeled their appearance in ACS images of M101 using version 6.0 of
TinyTIM \citep{ttim}.
M31 is about 10 times closer than M101, so the STIS images were 
input to TinyTIM as 10-times oversampled images, then convolved
with the HST/ACS point spread function using TinyTIM's ``scene-generation''
facility. Appropriate PSFs for the three ACS filters were used, but no
attempt was made to model the color-dependence of cluster appearance;
the wide bandpass of the STIS 50CCD filter made this unnecessary.
We computed PSFs and modeled artificial M101 clusters on a grid of 144 positions
on the ACS WFC detectors.

The artificial clusters were used first to define the cluster-selection 
procedure. We chose $V$ magnitudes randomly from a Gaussian with mean 22.0 and 
dispersion 1.2, $B-V$ and $V-I$ colors from Gaussians with means of 0.72 and 0.96
and dispersion 1.2, and reddening from a uniform distribution with $0<E(B-V)<0.2$.
The clusters were then added into the ACS images of field 1
(the innermost pointing)
on their PSF grid positions. After experimenting with object 
detection and measurement schemes, we decided to use SExtractor
\citep{ba96} for object detection and measurement,\footnote{ 
SExtractor parameters were set to the following values:
SEEING\_FWHM 0\farcs1, DETECT\_MINAREA=4, DETECT\_THRESH=1.5,
and the images were convolved with a $5\times5$ pixel Gaussian
filter before object detection.}
followed by FWHM measurements with the {\sc iraf} task {\sc apphot.radprof}.
Final cluster candidate selection criteria were:
objects which appear in all three of $B$, $V$, and $I$,
with MAG\_BEST$<23.5$, CLASS\_STAR$<0.05$, and ELLIP$<0.5$ as measured
on the $I$ images, and $2.2 < {\rm FWHM} < 6$~pixels as measured by
{\sc radprof}. Adding the magnitude criterion ensured that objects
would be bright enough for shape measurements and star/galaxy
classifications to be meaningful. Adding the FWHM criterion removed many 
extended sources which appeared to be galaxies,
reducing the number of candidates to a manageable level
while retaining most of the inserted artificial clusters.
The upper limit on FWHM of 6 pixels corresponds to an effective
radius $R_e=14.4$~pc (see Section~\ref{sec:sizedist}) at the distance of M101.

The actual cluster-finding procedure consisted of applying the above
criteria to each of the ten fields. Before finding clusters, we inserted
20 artificial clusters in each field. The same scheme described above
was used, except that the cluster locations were chosen randomly
and the cluster image used was the one with the closest PSF location.
After doing object detection and applying the cluster criteria,
the remaining candidates were visually examined to remove objects which
appeared to be background galaxies or blended stars. The visual inspection
imparts some subjectivity to the selection procedure; however
it proved to be necessary since we were unable to define an
automatic selection procedure which was both complete (picking
out most of the artificial clusters) and reliable (not picking too many 
obvious contaminants). The fraction of candidates removed by
the visual selection ranged from almost 50\% in the central field
to about 30\% in the outer fields.

\begin{figure}
\plotone{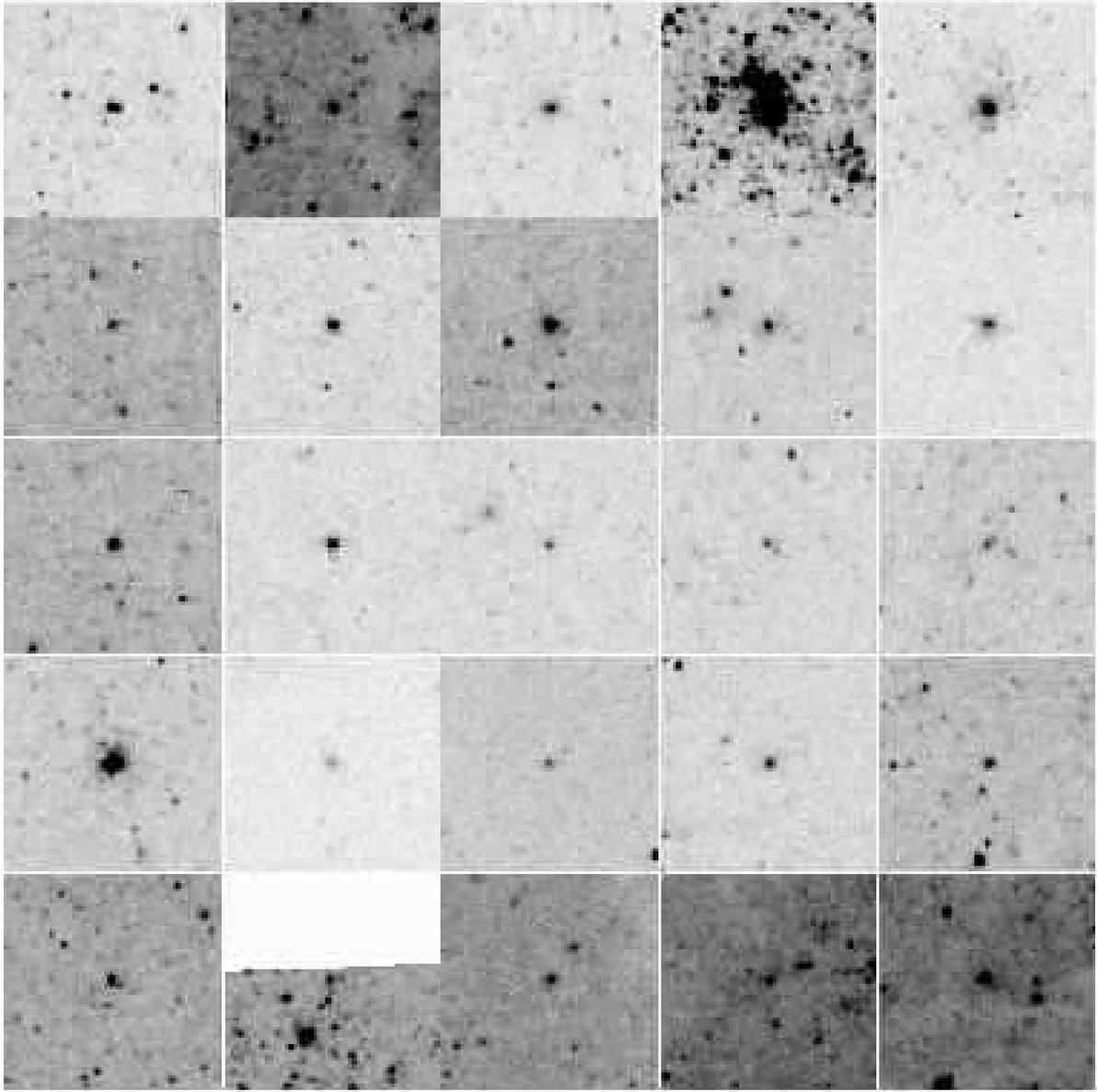}
\caption{Images of 25 randomly-chosen M101 cluster candidates. Each image is 
$5\farcs05 \times 5\farcs05$ (101 ACS WFC pixels). \label{samp-clust}}
\end{figure}

The completeness of the selection procedure is not well-defined
because of the subjective visual inspection. However, it can be estimated
by considering the fraction of artificial clusters which survived
the selection procedure. This ranged from 70\% in the central field to
90\% in several of the outer fields, with an average of 83\%.
A few clusters were missed because they were inserted at magnitudes
beyond the cutoff; most of the rest were deleted by the visual
inspection procedure, usually because they were not quite compact
enough to look like distinct objects. 
The final result of the cluster selection procedure was a list of 2920
candidates.
Images of a few candidates chosen at random from the final list
are shown in Figure~\ref{samp-clust}. The cluster candidates are 
reasonably round and appear resolved compared to nearby stars;
further analysis of their light distribution is in Section~\ref{sec:sizedist}.

Photometry of the cluster candidates was extracted from the
SExtractor output. Total magnitudes were measured with the SExtractor 
MAG\_AUTO estimator. Colors were measured
using aperture magnitudes in 2.5~pixel radius apertures. 
The aperture
magnitudes were  corrected to large apertures using aperture
corrections derived from the clusters themselves; since the clusters
are slightly resolved, corrections based on the instrument
point spread function alone are inappropriate. Corrections
used were $-0.905$, $-0.941$ , and $-1.055$ magnitudes 
in the F435W, F555W, and F814W filters, respectively. 
Transformations to the standard Johnson-Cousins system were made
using the equations and parameters in \citet{acs_cal}; no
corrections for charge transfer efficiency were made, but 
these are expected to be small since the observations were made
early in the life of ACS.

We used the H~I column density map given by \citet{kdk03} and 
a dust-to-gas ratio of $E(B-V) = 1.72\times10^{-22}$N(HI+H$_2$) \citep{boh78}
to de-redden the colors of the M101 cluster candidates. The column density
map reflects the total absorbing column in the M101 disk, so this correction
will over-estimate the reddening for objects in front of the disk plane (i.e.,
some  globular clusters.) Without further information, there is no way to tell 
which clusters are in front of the disk and which are behind, and the
reddening corrections are mostly small (the median value in the area
covered by the ACS images is $E(B-V) = 0.05$, with a maximum value of 0.26).
Galactic extinction in the direction of M101 is very low
\citep[$E(B-V)<0.01$;][]{cwl04} so we do not correct for it.

We can compare the results of our selection with the study of \citet[hereafter BKS]{bks96}.
These authors performed a visual search for clusters in one HST/WFPC2 field
which is contained within our central ACS pointing.\footnote{We discovered
that the coordinates given for BKS' clusters 30--43 do not fall on any
obvious objects in the ACS image; it appears that these objects fall on
the WF4 chip, but their RA and Dec were computed using the world coordinate
system for the PC1 image.
With the correct transformations all except two of the coordinate pairs
fall on obvious cluster candidates in the WF4 image.}
Their 43 cluster candidates have a median $V$ magnitude
of 21.5. Two of their candidates fall in the gap between the 
ACS detectors, and two are not identifiable on the WFPC2 image.
In the same area searched by BKS, 
our cluster selection procedure picked 232 candidates with a 
median magnitude of $V=22.8$.
Twenty-two of the remaining 39 BKS candidates (56\%)
are in our candidate list. Visual examination of the 17 
BKS clusters not in our list shows them to have a range of 
morphologies, but they are generally less round and less isolated
than the clusters on our list (see also Figure~\ref{m101-detail}). 
As indicated by the median magnitudes,
our candidates are generally fainter than those of BKS.
Comparison of photometry for objects in common shows a small
offset: median differences between their photometry and
ours are $\Delta V = 0.16\pm 0.04$, $\Delta(B-V) = -0.14\pm 0.03$, 
and $\Delta(V-I) = +0.08\pm 0.03$. BKS state that their
magnitudes are uncertain by $\geq 0.1$~mag, so we do not believe
the offset indicates a serious problem in our photometry.

\section{Analysis}

Star cluster system studies have generally focused on two separate
populations: the old, globular clusters or the young, massive clusters.
Elliptical galaxies, of course, have only the first type of cluster,
while the clusters studied in  galaxy mergers exemplify the second.
The color distribution of the M101 cluster candidates is expected
to reflect their age distribution: while colors of globular clusters
are often used as indicators of metallicity \citep[e.g.][]{kw98}, many of the
M101 candidates are far too blue to be old, metal-poor clusters.
In the following analysis, we will use color as a simple observational
distinction between older and younger clusters, to ease comparison
with other star cluster studies. $B-V=0.45$ is the red limit
used by \citet{lr99} for their studies of `Young Massive Clusters',
and is also close to the blue limit of the Milky Way globular cluster
$(B-V)_0$ distribution, so we adopt $(B-V)_0=0.45$ as the dividing line between
old and young clusters. It should be remembered, however, that the
age distribution of star clusters in M101 is not necessarily 
bimodal, and this division is somewhat arbitrary.
More precise age estimates will be enabled with the use of
UV and/or H$\alpha$ imaging; such data already exist for a portion of the
observed field and will be utilized in a future paper.

\begin{figure}
\plotone{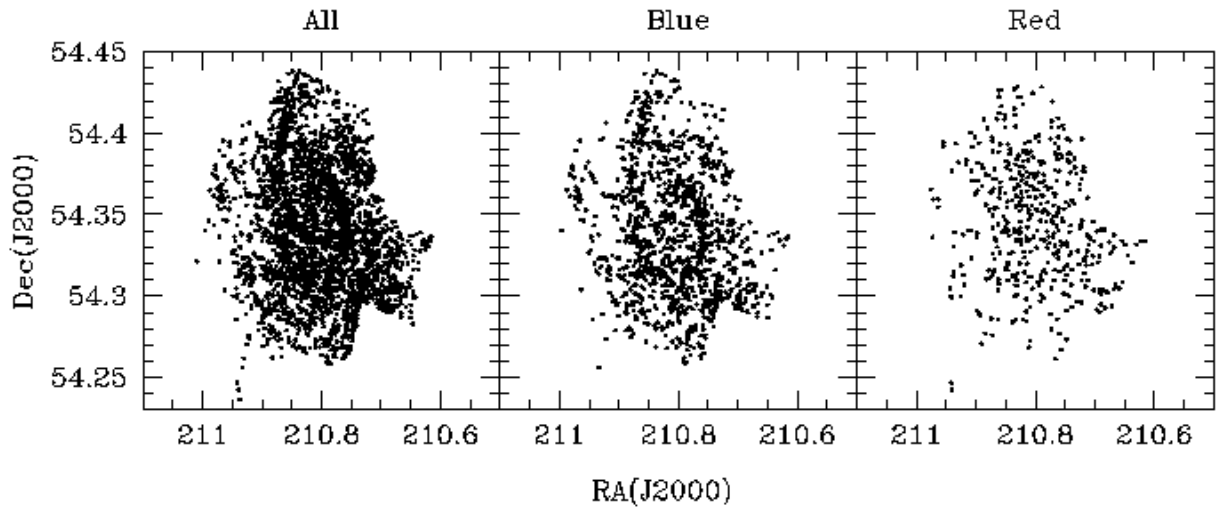}
\caption{Spatial distribution of M101 cluster candidates. The left panel
contains all 2920 candidates, while the right two panels show the red and blue
clusters separately. They contain only clusters with $V<23$, for which the sample 
is not biased by color.
\label{color-spat}}
\end{figure}

Of the 2920 star cluster candidates in our sample, 1877 (65\%) are
blue, with $(B-V)_0<0.45$. The effect of our $I=23.5$ magnitude limit is 
such that our cluster sample contains faint clusters only if they are red. 
The bluest clusters in our sample have $V-I \approx -0.5$, so the sample is 
not color-biased above $V=23$ $(M_V=-6.1)$. To this limit, there are
1715 cluster candidates of which 1260 (73\%) are blue.
Figure~\ref{color-spat} shows the spatial distribution of the
M101 candidates, both the complete sample and the bright ($V<23$) red and blue subsamples.
The spiral pattern of the galaxy is more apparent in the
blue clusters. This is consistent with them being younger and associated
with star formation in the spiral arms, and indeed the bluest clusters
($(B-V)<0.2$) trace the arms even more clearly.

\subsection{Color and luminosity distributions}

\begin{figure}
\plotone{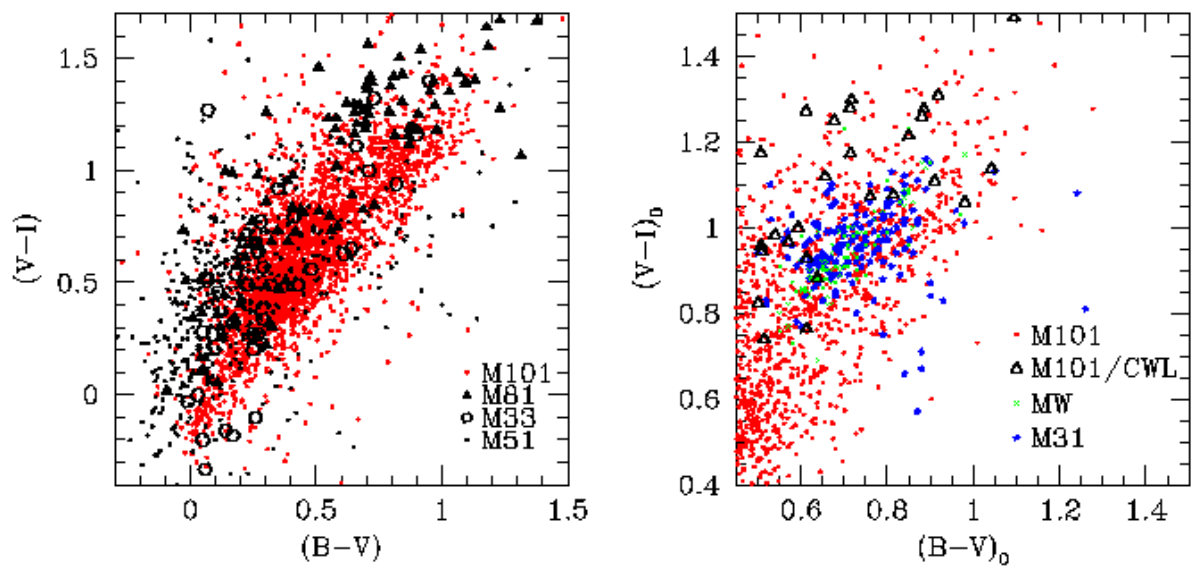}
\caption{Two-color plot for (candidate) clusters in spiral galaxies. 
Left: observed colors (not de-reddened)
for M101 cluster candidates from this work (small red dots), M51 cluster candidates
\citep[small black dots;][]{bik03}, M81 cluster candidates \citep[filled triangles;][]{cft1},
and M33 cluster candidates \citep[open circles;][]{m98}. 
Right: color distribution for red clusters, de-reddened. M101 cluster candidates
from this work (small red dots), M101 globular cluster candidates \citep[open triangles;][]{cwl04},
Milky Way globular clusters \citep[green crosses;][]{h96}, and M31 globular clusters
\citep[small blue stars;][]{b00}.
\label{2color}}
\end{figure}

Observed color distributions for the candidates are shown in Figure~\ref{2color}.
The left panel of this figure compares the colors for M101 clusters to those
in other nearby spirals: M81 \citep{cft1}, M33 \citep{m98}, and
M51 \citep{bik03}; these colors have not been corrected for reddening. 
The color distribution of M101 candidates is most similar to that of the M33
candidates, which is unsurprising given that the two galaxies are the same
Hubble type and have a similar specific star formation rate.
M101 has fewer blue clusters than M51 and fewer red clusters
than M81. This is broadly consistent with a picture in which the
proportion of blue to red clusters reflects the recent star formation rate
(M51 might be expected to have enhanced star formation due to encounters
with its companion), and the number of red clusters is proportional to
the galaxy bulge mass (which is larger for M81, an earlier-type galaxy than M101).
\citet{cft2} interpret the separation of M81 cluster
colors into two groups (at $B-V\approx0.5$ and $V-I\approx1.0$)
as a separation in age. No such clear separation is 
apparent in the M101 cluster candidates.

In the right panel of Figure~\ref{2color} we compare colors of red M101 cluster candidates
to those of globular clusters in the Milky Way \citep{h96}, M31 \citep{b00}, 
and globular cluster candidates in M101 itself from \citet{cwl04}.
The median colors of M101 cluster candidates are clearly consistent with 
those of the globulars in the other galaxies, indicating that we have
indeed detected a population of old globular clusters in M101.
The larger scatter of the M101 colors presumably reflects the larger 
photometric or reddening-correction errors.

\begin{figure}
\plotone{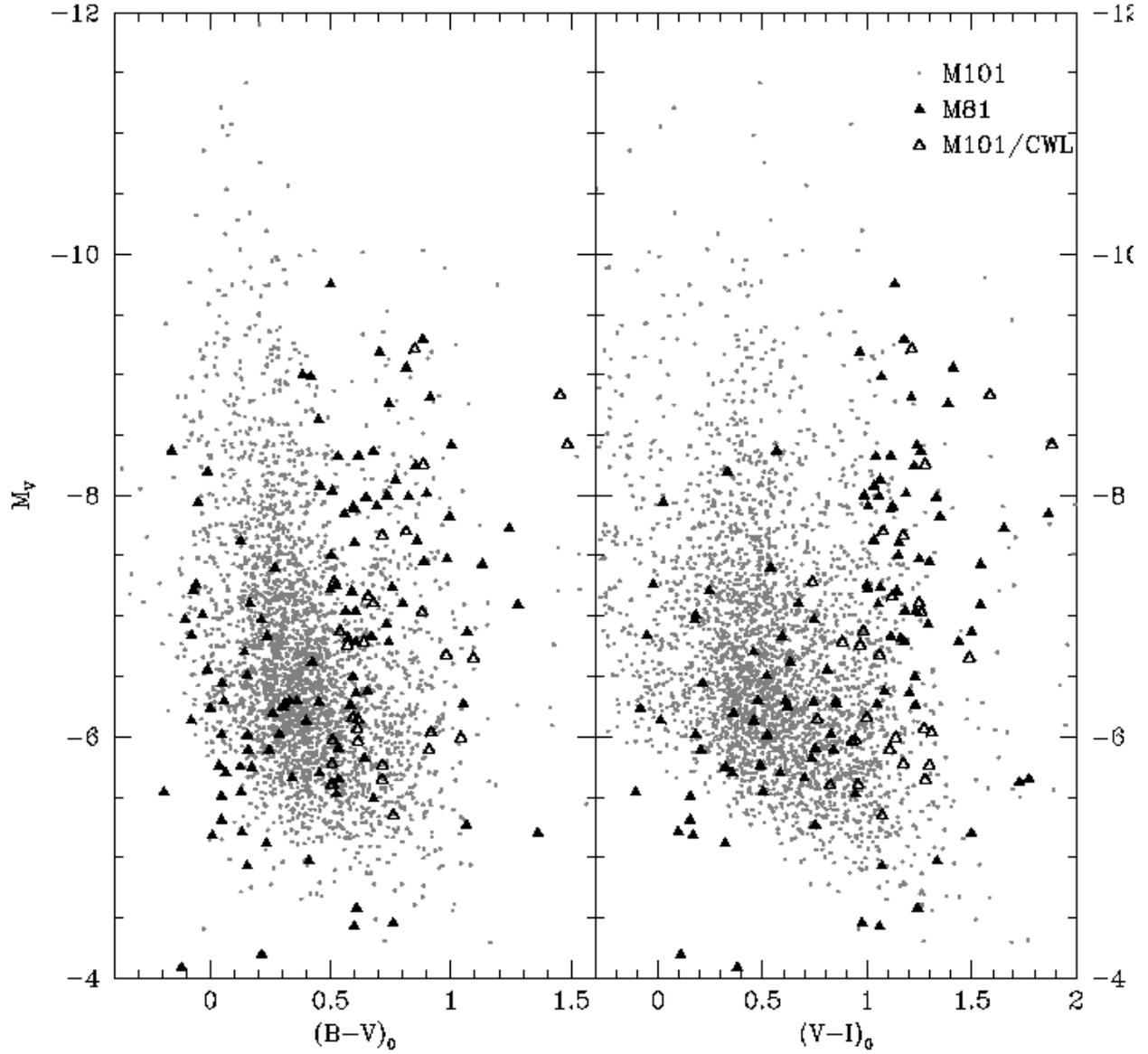}
\caption{Color-magnitude diagrams for star clusters in M101 and M81.
Symbols: M101 cluster candidates from this work (small grey dots),
M81 cluster candidates \citep[filled triangles;][]{cft1},
M101 globular cluster candidates \citep[open triangles;][]{cwl04}.
\label{cmd}}
\end{figure}

Figure~\ref{cmd} compares the joint distribution of colors and 
absolute magnitudes for M101, M81, and the M101 clusters found by \citet{cwl04}.
\citet{cwl04} were specifically attempting to
select old, globular clusters in M101, while \citet{cft1} were searching only
for `compact' star clusters in M81 with no selection on age.
The effect of our $I$-band magnitude cut is particularly
clear in the right-hand panel of this figure: our cluster sample is
missing faint, blue clusters.
It is clear that the excess of red clusters in M81 compared to M101 is
mostly at brighter cluster luminosities: brighter than $M_V=-7.4$ (the expected
peak of the globular cluster luminosity function), the M81 clusters are mostly
red, while the M101 clusters are mostly blue. 
We also confirm the detection by \citet{cwl04} of
a number of faint, red clusters in M101. Some of these could be
faint background galaxies, especially ellipticals
or the bulges of spirals whose disks are too faint to observe.
Such contamination is unlikely to account for all of the faint red
clusters, whose nature is discussed further in Section~\ref{sec:spatdist}.

\begin{figure}
\plotone{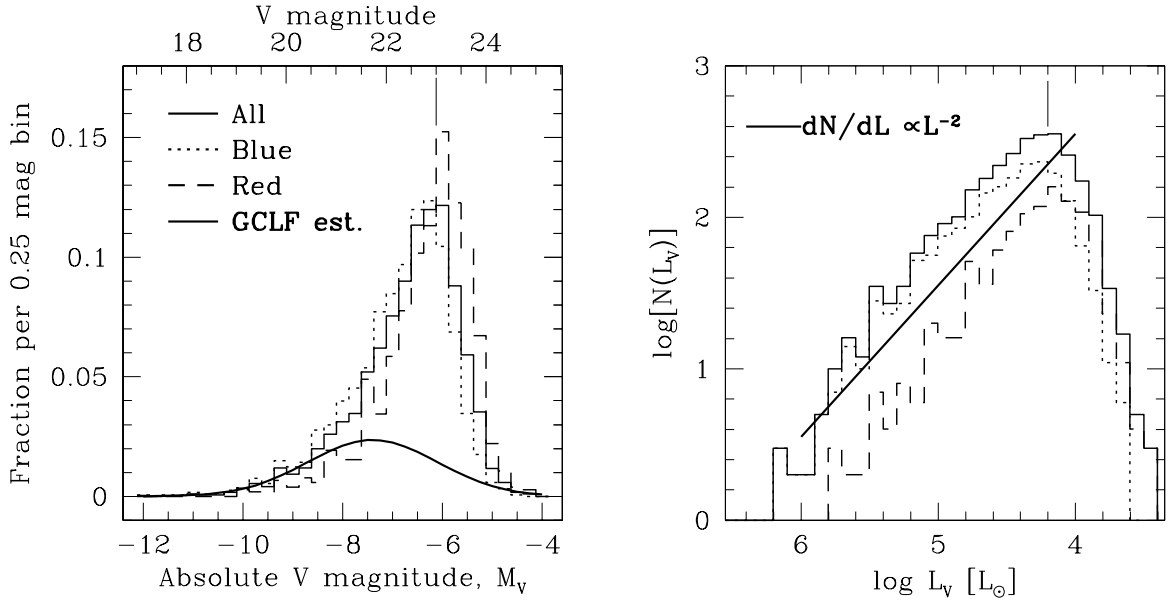}
\caption{(Left) M101 star cluster candidate magnitude distributions: all clusters
(solid histogram), blue clusters (dotted histogram), red clusters
(short-dashed histogram). The heavy smooth curve shows the
expected globular cluster luminosity function, scaled to the number of
red clusters with $M_V<-7.4$. (Right) M101 star cluster candidate luminosity 
distributions. Line types as in left panel, except that the heavy solid line 
illustrates a power-law distribution $dN(L)/dL \propto L^{-2}$ (not a 
fit to the data). In both panels, the vertical line indicates the $V=23$ limit
below which the blue cluster sample is incomplete.
\label{lum-dist}}
\end{figure}

The cluster candidate luminosity distribution is shown in Figure~\ref{lum-dist},
for the full sample and the red and blue subsamples. 
The strong fall-off at $V>23$ ($M_V>-6.1$) is due to our imposed magnitude limit (see above).
Brighter than this limit, we find that the blue clusters are about
0.25~mag brighter in the median than the red clusters. From population
synthesis models, such a difference is consistent with the blue clusters being 
younger, by a few Gyr if both populations are $\gtrsim 8$~Gyr old, or
less if the clusters are younger. Inferred ages are of course strongly dependent on
additional factors such as metallicity, reddening, and initial mass function.
As discussed below, there are many more red clusters than the expected number of
globular clusters for a galaxy of M101's luminosity: the smooth curve
in the left panel of the figure shows a `standard' globular cluster luminosity function
(a Gaussian with mean $M_V=-7.4$ and standard deviation $\sigma=1.3$)
scaled to the number of clusters with $M_V<-7.4$ (also see Section~\ref{sec:spatdist}).
The bright end of the LF for the full sample is consistent with the power-law 
distribution of luminosity $dN(L)/dL \propto L^{-2}$ ($N(L) \propto L^{-1}$), 
similar to the distributions seen for young clusters in mergers \citep{whi99,mwsf97}
and {\it HST}-identified (not necessarily young) clusters in other spirals \citep{lar02b},
as well as the bright end of the GCLF for the Milky Way and M31 \citep{hp94,mcl94}.

\subsection{Cluster populations and spatial distribution\label{sec:spatdist}}

Do the blue and red  subsamples of M101 correspond to 
`young' and `globular'  cluster groups? One way to find out is to compare
the number of clusters per unit galaxy luminosity in M101 to values
for other galaxies. The total magnitudes of M101 as given by \citet{rc3} 
are $B_T=8.31$, $V_T=7.86$, corresponding to luminosities of $M_B=-20.82$, $M_V=-21.27$.
The `young massive cluster' (YMC) criterion used by 
\citet{lr99} in their analysis of such clusters
in nearby spirals includes both color ($B-V<0.45$) and faint magnitude limits, $M_V<-8.5$
(the latter to avoid contamination by individual massive stars). 140 of our candidates
meet these criteria, giving a value of $T_N=N_{\rm YMC} \times 10^{0.4(M_B+15)}=0.66$.
This is a fairly typical value for spirals in the \citet{lr99} survey: although 
M101 has more YMCs than all but one of the \citet{lr99} galaxies, it is
also more luminous than any of them. 
Only the brightest M101 blue clusters are YMCs. To compare the overall cluster 
population to other spirals we can use ${\Sigma}_{\rm cl}^{-8}$, the number density of clusters per
kpc$^2$ per 1 magnitude bin at $M_V=-8$ \citep[defined by][]{lar02b}.
M101 has ${\Sigma}_{\rm cl}^{-8}=1.0$, which is quite typical of spirals.

Globular cluster populations in different galaxies are usually compared
by means of the specific frequency, defined as $S_N= N_{\rm GC} 10^{0.4(M_V+15)}$.
If we assume that all 1043 red objects are globular clusters,
the corresponding globular cluster specific frequency of M101 is $S_N=3.0$.
This value is quite high compared to other 
spiral galaxies, which typically have values in the range 0.5--1.0 \citep{pmb03, cwl04},
so assuming M101 to be a typical spiral would tend to support the argument
that the faint red clusters are not globulars.
As another estimate of $S_N$ in M101, we can {\em assume} that the M101
globular clusters have the usual Gaussian luminosity function with a peak
at $M_V=-7.4$ \citep[following][]{cwl04}, that our cluster selection
is complete to this magnitude limit, 
and that all red clusters brighter than this limit are true
globulars. There are 160 such clusters, so if the luminosity function
assumption is correct, $S_N=(320/1043)\times3.0 = 0.9$, a much more typical
value for spiral galaxies, although still higher than the $0.5\pm0.2$ found
for late-type spirals by \citet{cwl04}. These authors found $S_N= 0.4\pm0.1$
for M101 specifically; however, this is an extrapolation from a small
number of detected globulars (29).

Are M101's `excess' faint red clusters low-mass old globulars
which (unlike in earlier-type spirals) have not been destroyed by
dynamical effects? Or are they instead reddened young clusters?
At least some are likely to be true old clusters, since \citet{cwl04} 
found a similar population after using $U$-band information to eliminate 
younger clusters from their sample. However, since we use only $BVI$ colors, we 
suspect that the majority of our faint red clusters are likely to be reddened
younger clusters.  The HI map upon which the column density map is based 
has a resolution of $\sim15\arcsec$ (the beam was $13\arcsec\times16\arcsec$)
while the resolution of the 12m CO map was $\sim55\arcsec$.
Since young clusters are often associated with current
star-formation and localized concentrations of neutral and molecular gas, the
low spatial resolution of the column density map could
cause the reddening of younger clusters to be underestimated.
If we have underestimated the reddening of some clusters, their absolute
magnitudes would also be underestimated: correcting for this effect would
then imply that the faint red clusters were intrinsically brighter
than the faint blue clusters. This is a selection effect rather than a physical
difference, however: more heavily-reddened
clusters must be intrinsically brighter to be observable \citep[for a discussion
of this effect in M31, see][]{bpb02}. Given the similarity of the luminosity
functions of faint red and blue clusters (see previous section), we believe it
more likely that the faint red clusters are mostly reddened younger clusters.

\begin{figure}
\plotone{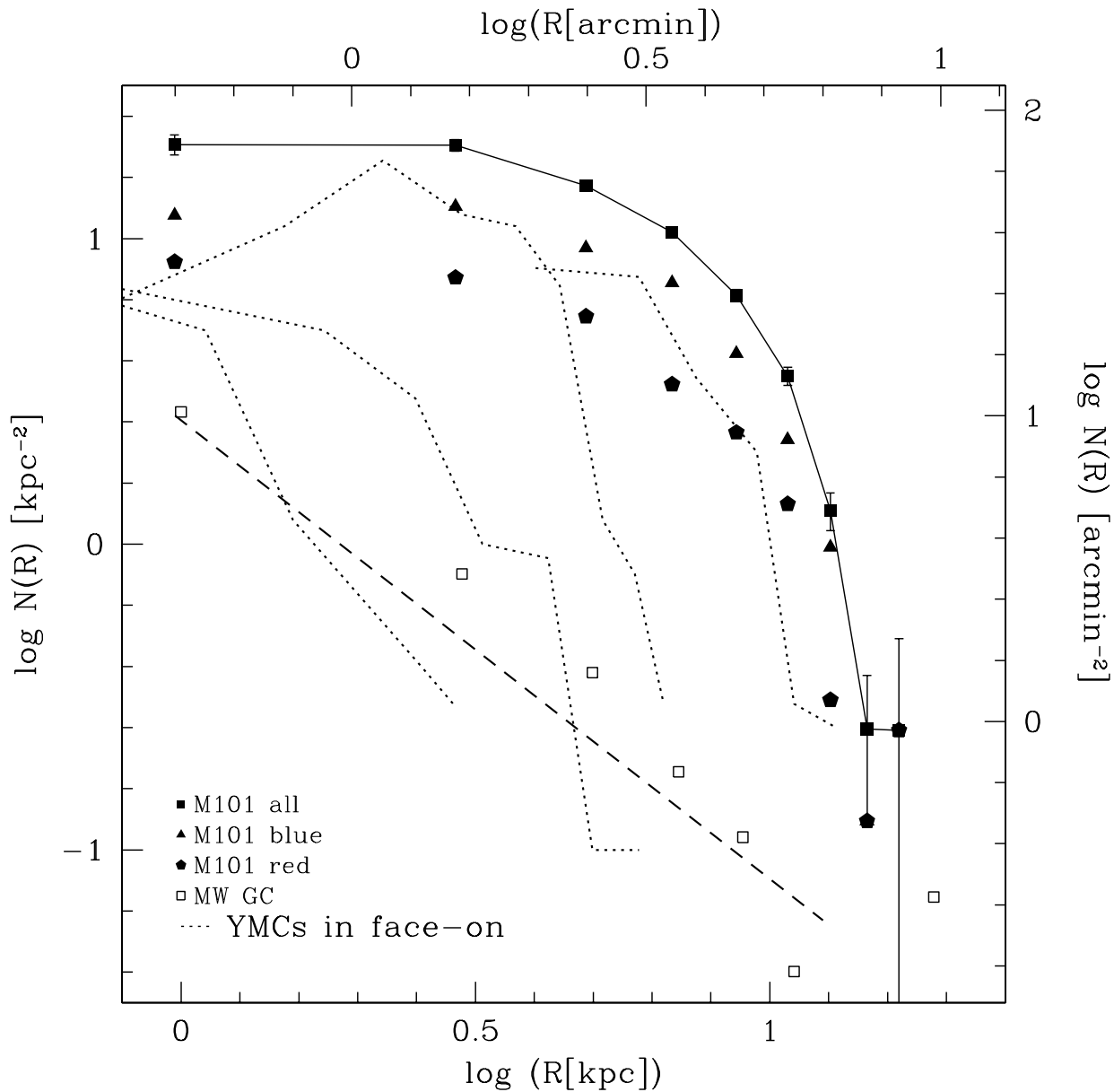}
\caption{Cluster candidate surface density distribution for all M101 clusters (filled squares),
M101 blue clusters (triangles), M101 red clusters (pentagons), and
Milky Way globulars (open squares)
Dotted lines: cluster surface density distributions in NGC~1156, NGC~1313, NGC~2997
and NGC~5236 from \citet{lr99}, with arbitrary vertical normalization.
The axes labeled in kpc apply to all galaxies, while the labels 
in arcminutes apply only to M101.
The dashed line illustrates a power-law distribution $N(R)\propto R^{-1.5}$; it is
not a fit to the data for any galaxy.
\label{surf-dens}}
\end{figure}

Next we consider the spatial distribution of the M101 clusters.
Using the results of the artificial cluster experiments described above, we 
made a `completeness map', which accounts for both the spatially-varying
completeness and the area on the sky actually observed as a function
of distance from the center of M101.
As Figure~\ref{m101-fields} shows, our 
observations cover most of the central 5\arcmin\ (9.75 kpc) radius
of the galaxy, but only a small fraction of the region
between 5 and 10 arcmin. 
In a set of annular bins, we counted the observed number of clusters and
divided by the completeness-corrected, observed area of the annuli.
The cluster candidate surface density 
distribution is shown in Figure~\ref{surf-dens}. For comparison
we also show the surface density distributions of
young massive clusters in four spirals from \citet[their Figure 8]{lr99},
and of Milky Way globular clusters, using positions from \citet{h96} 
projected on to the plane of the Galaxy.

The overall normalization of the cluster surface density distribution in M101 is
higher than in the Milky Way. This is not unexpected since
the latter includes only globular clusters.
However, the shape of the  distribution is also different in M101: while the Milky Way
globulars show a power-law distribution $n(R)\propto R^{-\alpha}, \alpha\sim-1.5$,
the M101 and young massive cluster distributions look much more like King profiles, with
a constant density core and rapid fall-off beyond.
This is consistent with the M101 and other young massive clusters mostly 
belonging to the galaxy disks rather than the more extended halo
usually identified with globular clusters.

\subsection{Size distribution\label{sec:sizedist}}

\begin{figure}
\plotone{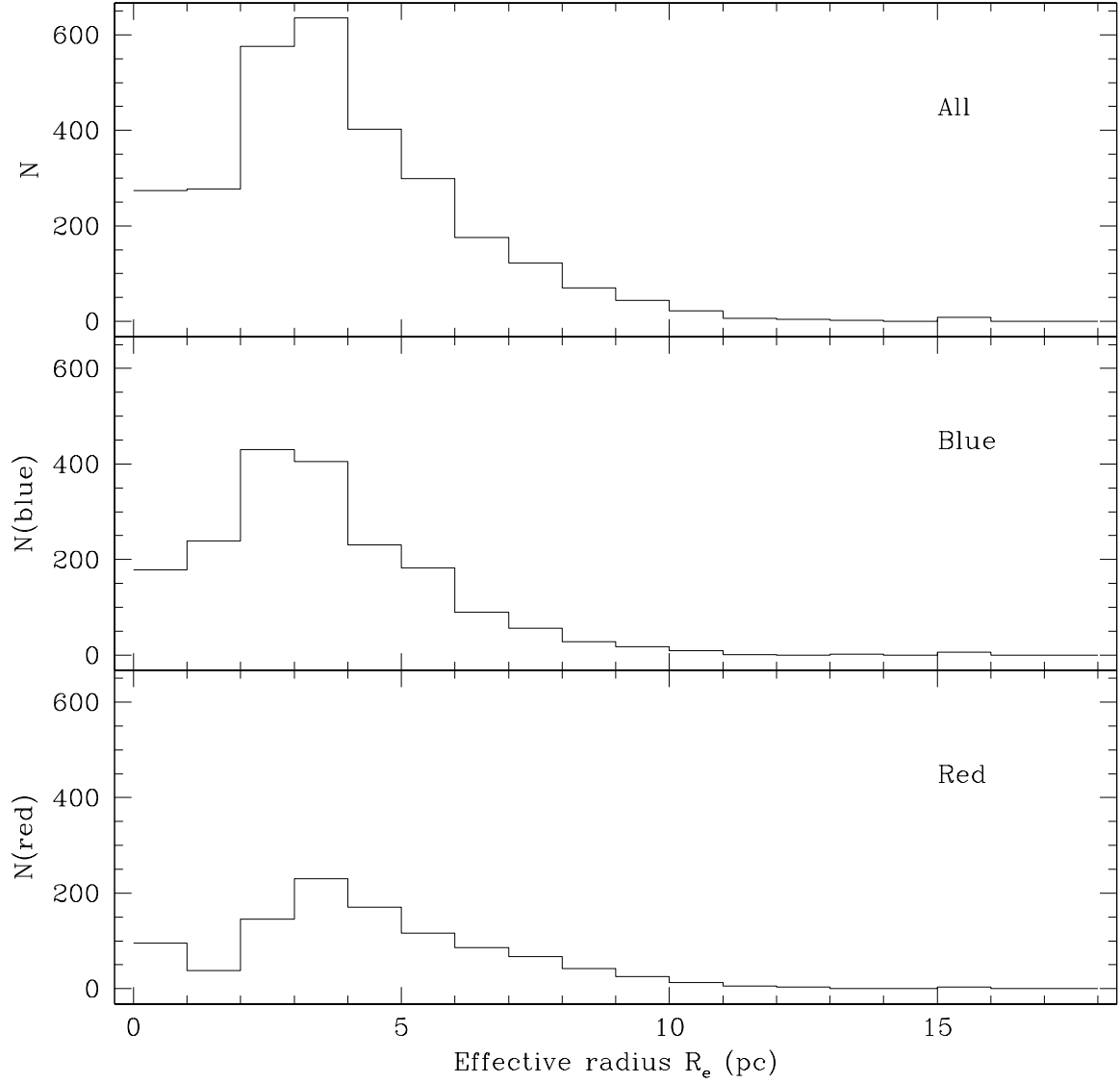}
\caption{Cluster candidate effective radius distribution for all M101 cluster candidates
(top), blue candidates (center), and red candidates (bottom).\label{size-dist}}
\end{figure}

We measured cluster candidate sizes using the {\sc ishape} program 
\citep{lar99}. This software fits a number of PSF-convolved model light 
distributions to the images of individual objects. We fit a King model 
with concentration index $c=30$, for the most direct comparison
with the results of \citet{lar99}, although the derived effective radius 
is not particularly sensitive to the model chosen. We fit only circular
models, as experiments showed that using elliptical light models
did not significantly change the measured sizes but did increase the
computing time by a factor of $\sim 3$. The distribution of cluster candidate
effective radii $R_e$ (the radius which contains half of the integrated
light) is shown in Figure~\ref{size-dist}. The median $R_e$ was 2.2 ACS
pixels, or 3.5~pc, which confirms that our selection procedure did indeed
select mostly slightly extended objects. Fewer than 10\% of our candidates were
flagged as `possibly stellar' by {\sc ishape}.

There is a difference in the size distributions of red and blue clusters.
The red clusters, with a median $R_e=4.1$~pc, are slightly larger
than the blue clusters, whose median size is $3.2$~pc. Dynamical studies 
of star cluster evolution \citep[e.g. ][]{go97} show that large clusters
are more easily destroyed, so we expected that the red (older) clusters
would be slightly smaller on average than the blue (younger).
However, since the clusters are only marginally resolved, 
the difference between the two groups is not of major significance and
could reflect contamination of the red
cluster sample by younger clusters or galaxies. The range of blue cluster candidate sizes 
is similar to the range found by \citet{lar99}
for `young massive clusters' and is also comparable to the median half-light radius 
for Milky Way globular clusters $R_e=3.2$~pc \citep{h96}.

\section{Summary}

We have presented the results from an initial search for star clusters
in Hubble Space Telescope/Advanced Camera for Surveys images of the
nearby giant spiral M101. Defining a reliable sample of cluster candidates
in these complex, crowded images proved to be a challenge.
Our final cluster candidate list contains nearly 3000 
objects of which a majority are blue and associated with the galaxy's spiral arms.
These cluster candidates have color distributions similar to 
candidates and confirmed clusters in other spirals including the Milky Way, M31, M81,
M33, and M51. M101 has fewer bright, red clusters than the earlier-type spiral
M81, but many more faint red clusters \citep[as originally noted by][]{cwl04}.
If all of these faint red clusters were globulars, M101 would have a much
higher globular cluster specific frequency and a much different 
globular cluster luminosity function from other galaxies. We suggest that
many of the red M101 clusters are likely to be reddened young clusters, 
which is supported by the consistency of the overall cluster luminosity
distribution and fiducial cluster density with those found in other spirals
by \citet{lar02b}. The spatial distribution of the M101 clusters shows a
relatively flat core and a steep drop-off at larger galactocentric distances,
suggesting that the clusters are associated with the galaxy disk rather than its halo.
The size distributions of the M101 cluster candidates show them to be slightly
resolved, as expected, and are consistent with cluster sizes measured in other galaxies.

\acknowledgments
Support for program GO-9490 was provided by NASA through a grant from the Space
Telescope Science Institute, which is operated by the Association of Universities
for Research in Astronomy, Inc., under NASA contract NAS 5-26555.

\begin{thebibliography}{}

\bibitem[{Barmby(2003)}]{pmb03}
Barmby, P. 2003, in Extragalactic Globular Cluster Systems, ed.
  M.~Kissler-Patig (Berlin: Springer-Verlag), 143

\bibitem[Barmby {et~al.}(2000)]{b00}
Barmby, P., Huchra, J.~P., Brodie, J.~P., Forbes, D.~A., Schroder, L.~L., \&
  Grillmair, C.~J. 2000, \aj, 119, 727

\bibitem[{{Barmby} {et~al.}(2002){Barmby}, {Perrett}, \& {Bridges}}]{bpb02}
{Barmby}, P., {Perrett}, K.~M., \& {Bridges}, T.~J. 2002, \mnras, 329, 461

\bibitem[{Bertin} \& {Arnouts}(1996)]{ba96}
{Bertin}, E. \& {Arnouts}, S. 1996, \aaps, 117, 393

\bibitem[{Bica} {et~al.}(2003)]{bica03}
{Bica}, E., {Dutra}, C.~M., {Soares}, J., \& {Barbuy}, B. 2003, \aap, 404, 223

\bibitem[{Bik} {et~al.}(2003)]{bik03}
{Bik}, A., {Lamers}, H.~J.~G.~L.~M., {Bastian}, N., {Panagia}, N., \&
  {Romaniello}, M. 2003, \aap, 397, 473

\bibitem[{Bohlin {et~al.}(1978)Bohlin, Savage, \& Drake}]{boh78}
Bohlin, R., Savage, B., \& Drake, J. 1978, \apj, 224, 132

\bibitem[{Bresolin {et~al.}(1996)Bresolin, Kennicutt, \& Stetson}]{bks96}
Bresolin, F., Kennicutt, R.~C., \& Stetson, P.~B. 1996, \aj, 112, 1009

\bibitem[{{Chandar} {et~al.}(2001a){Chandar}, {Ford}, \& {Tsvetanov}}]{cft1}
{Chandar}, R., {Ford}, H.~C., \& {Tsvetanov}, Z. 2001a, \aj, 122,
  1330

\bibitem[{{Chandar} {et~al.}(2001b){Chandar}, {Tsvetanov}, \& {Ford}}]{cft2}
{Chandar}, R., {Tsvetanov}, Z., \& {Ford}, H.~C. 2001b, \aj, 122,
  1342

\bibitem[{{Chandar} {et~al.}(2004){Chandar}, {Whitmore}, \& {Lee}}]{cwl04}
{Chandar}, R., {Whitmore}, B., \& {Lee}, M.~G. 2004, \apj, 611, 220

\bibitem[{de Vaucouleurs} {et~al.}(1991)]{rc3}
{de Vaucouleurs}, G., {de Vaucouleurs}, A., {Corwin}, H.~G., {Buta}, R.~J.,
  {Paturel}, G., \& {Fouque}, P. 1991, {Third Reference Catalogue of Bright
  Galaxies (RC3)} (New York: Springer-Verlag)

\bibitem[{Ford {et~al.}(2002)}]{acs}
Ford, H. {et~al.} 2002, SPIE, 4854, 81

\bibitem[{{Freedman} {et~al.}(2001)}]{hst_kp}
{Freedman}, W.~L. {et~al.} 2001, \apj, 553, 47

\bibitem[{{Gnedin} \& {Ostriker}(1997)}]{go97}
{Gnedin}, O.~Y. \& {Ostriker}, J.~P. 1997, \apj, 474, 223

\bibitem[{Goudfrooij} {et~al.}(2003)]{pg03}
{Goudfrooij}, P., {Strader}, J., {Brenneman}, L., {Kissler-Patig}, M.,
  {Minniti}, D., \& {Huizinga}, J. 2003, \mnras, 343, 665

\bibitem[{Harris(1996)}]{h96}
Harris, W.~E. 1996, \aj, 112, 1487

\bibitem[{{Harris} \& {Pudritz}(1994)}]{hp94}
{Harris}, W.~E. \& {Pudritz}, R.~E. 1994, \apj, 429, 177

\bibitem[{{Kelson} {et~al.}(1996)}]{kel96}
{Kelson}, D.~D. {et~al.} 1996, \apj, 463, 26

\bibitem[{{Krist}(1995)}]{ttim}
{Krist}, J. 1995, in Astronomical Data Analysis Software and Systems IV, ed.
  R.~Shaw, H.~Payne, \& J.~Hayes, ASP Conf. Ser., 349

\bibitem[{{Kundu} \& {Whitmore}(1998)}]{kw98}
{Kundu}, A. \& {Whitmore}, B.~C. 1998, \aj, 116, 2841

\bibitem[{Kuntz} {et~al.}(2005)]{kdk05}
{Kuntz}, K.~D., {Gruendl}, R.~A., {Chu}, Y.-H., {Chen}, C.-H.~R., {Still}, M.,
  {Mukai}, K., \& {Mushotzky}, R.~F. 2005, \apjl, 620, L31

\bibitem[{Kuntz} {et~al.}(2003)]{kdk03}
{Kuntz}, K.~D., {Snowden}, S.~L., {Pence}, W.~D., \& {Mukai}, K. 2003, \apj,
  588, 264

\bibitem[{{Larsen}(1999)}]{lar99}
{Larsen}, S.~S. 1999, \aaps, 139, 393

\bibitem[{{Larsen}(2002)}]{lar02b}
{Larsen}, S.~S. 2002, \aj, 124, 1393

\bibitem[{{Larsen} \& {Richtler}(1999)}]{lr99}
{Larsen}, S.~S. \& {Richtler}, T. 1999, \aap, 345, 59

\bibitem[{{Lee} {et~al.}(2005){Lee}, {Chandar}, \& {Whitmore}}]{lcw05}
{Lee}, M.~G., {Chandar}, R., \& {Whitmore}, B.~C. 2005, \aj, 130, 2128

\bibitem[{{McLaughlin}(1994)}]{mcl94}
{McLaughlin}, D.~E. 1994, \pasp, 106, 47

\bibitem[{Miller} {et~al.}(1997)]{mwsf97}
{Miller}, B.~W., {Whitmore}, B.~C., {Schweizer}, F., \& {Fall}, S.~M. 1997,
  \aj, 114, 2381

\bibitem[Mochejska {et~al.}(1998)]{m98}
Mochejska, B.~J., Kaluzny, J., Krockenberger, M., Sasselov, D.~D., \& Stanek,
  K.~Z. 1998, Acta Astronomica, 48, 455

\bibitem[{Mukai} {et~al.}(2003)]{mukai03}
{Mukai}, K., {Pence}, W.~D., {Snowden}, S.~L., \& {Kuntz}, K.~D. 2003, \apj,
  582, 184

\bibitem[{{Shafter} {et~al.}(2000){Shafter}, {Ciardullo}, \&
  {Pritchet}}]{scp00}
{Shafter}, A.~W., {Ciardullo}, R., \& {Pritchet}, C.~J. 2000, \apj, 530, 193

\bibitem[{{Sirianni} {et~al.}(2005)}]{acs_cal}
{Sirianni}, M. {et~al.} 2005, \pasp, 117, 1049

\bibitem[{Stetson {et~al.}(1998)}]{ste98}
Stetson, P.~B. {et~al.} 1998, \apj, 508, 491

\bibitem[{Strader} {et~al.}(2005)]{str05}
{Strader}, J., {Brodie}, J.~P., {Cenarro}, A.~J., {Beasley}, M.~A., \&
  {Forbes}, D.~A. 2005, \aj, 130, 1315

\bibitem[{Whitmore} {et~al.}(1999)]{whi99}
{Whitmore}, B.~C., {Zhang}, Q., {Leitherer}, C., {Fall}, S.~M., {Schweizer},
  F., \& {Miller}, B.~W. 1999, \aj, 118, 1551

\end{thebibliography}
\end{document}